\documentstyle[12pt,epsfig]{article}

\begin{document}

\centerline{\bf\Large  Possibly stable configurations of
$\Theta^+_5$ in the flux-tube model}

\vspace{1cm}

\centerline{Zhao Hong-ming$^{1,2}$, Zeng Zhuo-quan$^{1,2}$,Shen
Peng-Nian$^{4,1,5,6}$, Ding Yi-Bing$^{2,4}$  and  Li
Xue-Qian$^{3,4,5}$}

\vspace{1cm}

1. Institute of High Energy Physics, P.O. Box 918-4, Beijing, 100049, China.

2. College of Physical Sciences, The Graduate School of the Chinese Academy
of Sciences, Beijing, 10049, China.

3. Department of Physics, Nankai University, Tianjin, 300071, China.

4. CCAST(World Lab.), P.O.Box 8730, Beijing 100080, China

5. Institute of Theoretical Physics, Chinese Academy of Sciences,
 P.O.Box 2735, Beijing 100080, China

6. Center of Theoretical Nuclear Physics, National Laboratory of
Heavy Ion Accelerator, Lanzhou 730000, China

\vspace{1cm}

\begin{center}
\begin{minipage}{12cm}
\noindent {\bf Abstract}

\vspace{1cm}

In this work, following the literature, we investigate the
semi-classical picture where various spatial configurations for
$\Theta^+_5$ were suggested by several authors. We use the flux
tube model where the effective Column coefficient and string
tension are extracted from the lattice QCD calculations, to obtain
the interactions among the constituent quarks, which is later to
be taken as an effective potential. It is reasonable to assume
that keeping the simple spatial shape, only the breathing mode is
considered, thus the whole system is much simplified and described
by only one variable. Substituting the potential into the
Schr\"odinger equation, we evaluate the binding energy of the
system. We find that each of the configuration has higher
symmetries and corresponds to a local minimum of energy and one of
them results in a mass which is smaller than the others. It is
interesting to notice that with such a very simplified
semi-classical picture the minimal value corresponds to a mass of
1530 to 1554 MeV for $\Theta^+_5$. The good agreement with the
LEPS' data may indicate that such simplified picture indeed
reflects physical reality in some sense.

\vspace{0.5cm}


\end{minipage}
\end{center}

\vspace{1cm}

Since the LEPS Collaboration \cite{Naka} at the SPring8  announced
the discovery of a sharp resonance with $M=1.54\pm0.01$ GeV and
$\Gamma<25$ MeV in the $\gamma+n\to K^+K^- $ process in 2003, it
has greatly attracted attentions of physicists. Because its
strangeness is +1, the observed $\Theta^+_5$ cannot be a normal
baryon with only three valence quarks, nowadays, it is suggested
that
the newly observed resonance is a five-quark state $\Theta_5^+~(uudd\bar{s}%
)$ \cite{Zhu}. If it is true, it will open up a new epoch in the
study of multi-quark states in hadron physics.

Since then, numerous experimental groups are involved in searching
and confirming such a state and its family members through various
reaction processes. Meanwhile, theorists have also been trying to
explain that experimental finding by employing various theoretical
models \cite{Zhu}. Even though remarkable progresses have been
made, unfortunately, a common viewpoint is still not reached yet,
in other words, the physical properties of the state, such as the
structure, the mass, the decay width, the parity and etc. are
still in puzzle.

The flux tube model \cite{Isg} is a  model based on the lattice
QCD, indeed it establishes a theoretical framework where mesons,
baryons and exotics are described in a unified way. Employing such
a model to study the five-quark state is obviously reasonable. In
fact, there have been several papers on this line. Okiharu {\it
{et al.}} \cite{Oki} suggested that for the lowest five-quark
state, the effective potential should be composed of the two-body
color-Coulomb interaction and a long-ranged confining potentia,
and the five constituents are connected by flux tubes into the
double Y-shape, shown in Fig.1(a). Alexandrou {\it {et al.}}
\cite{Alex} proposed an alternative structure that the quarks are
connected by flux tubes and the tubes intersect at three Steiner
points, shown in Fig.1(b). Song {\it {et al.}} \cite{Song} argued
that the state should have a tetrahedron (they called it as a
"diamond") structure depicted in Fig.1(c). Obviously, all the
structures are simplified semi-classical scenarios, but they seem
to reflect the physics reality in some sense. However, except a
simple variational calculation on the $\Theta^+_5$ spectrum where
a combined two-body interaction is employed\cite{Kana} has been
done, there are almost no quantitative calculations carried out
with the above mentioned models at the present moment. Therefore,
it would be interesting and significant to judge what type of the
flux tube structure corresponds to the stable one for $\Theta^+_5$
through a numerical evaluation of its mass. In the framework of
the flux tube model which is a phenomenological version of the
lattice QCD, the constituents in a hadron are connected by
color-flux tubes to construct a pattern with the minimal total
length of the tube, which provides a confining potential in the
hamiltonian. Meanwhile, the one-gluon-exchange (OGE) interaction
is also included in the effective hamiltonian to remedy the short
range interaction. Substituting the hamiltonian into the
Schr\"odinger equation and solving it, we obtain the binding
energy of the system and the mass of $\Theta_5^+$. Especially, we
may find whether the tetrahedral structure suggested by Song {\it
{et al.}} is the more stable one compared with the others.

In this letter, we suppose that the aforementioned semi-classical
flux-tube structures shown in Figs.1(a),(c),(d) and (e) are
relative stable, then estimate the mass of $\Theta^+_5$ with these
structures.

In the figures, the structure of Fig.1(c) is in a regular
tetrahedral form with four quarks sitting at four corners,
respectively, and the anti-strange quark staying at the center of
the tetrahedron. If twisting the triangle $\triangle_{CDE}$ around
the axis $ab$, where $a$ and $b$ are the midpoints of lines $AB$
and $CD$, respectively, by $\pi/2$, one obtains Fig.1(d).

According to Okiharu's suggestion, in our calculation, preserving
the shape of the rectangle, the flux tubes join the quarks and
anti-quark into the double Y form shown in Figs.1(a) and 1(e). The
difference between these two figures is that in Fig.1(a), the
quarks at two ends of the longer edges form color-$\bar 3$ states,
whereas in Fig.1(e), the two quarks at the ends of the shorter
edges constitute color-$\bar 3$ states.

\begin{figure}[htb]
\begin{center}
\vglue 2.cm

\parbox{.32\textwidth}{\epsfysize=4.5cm \epsffile{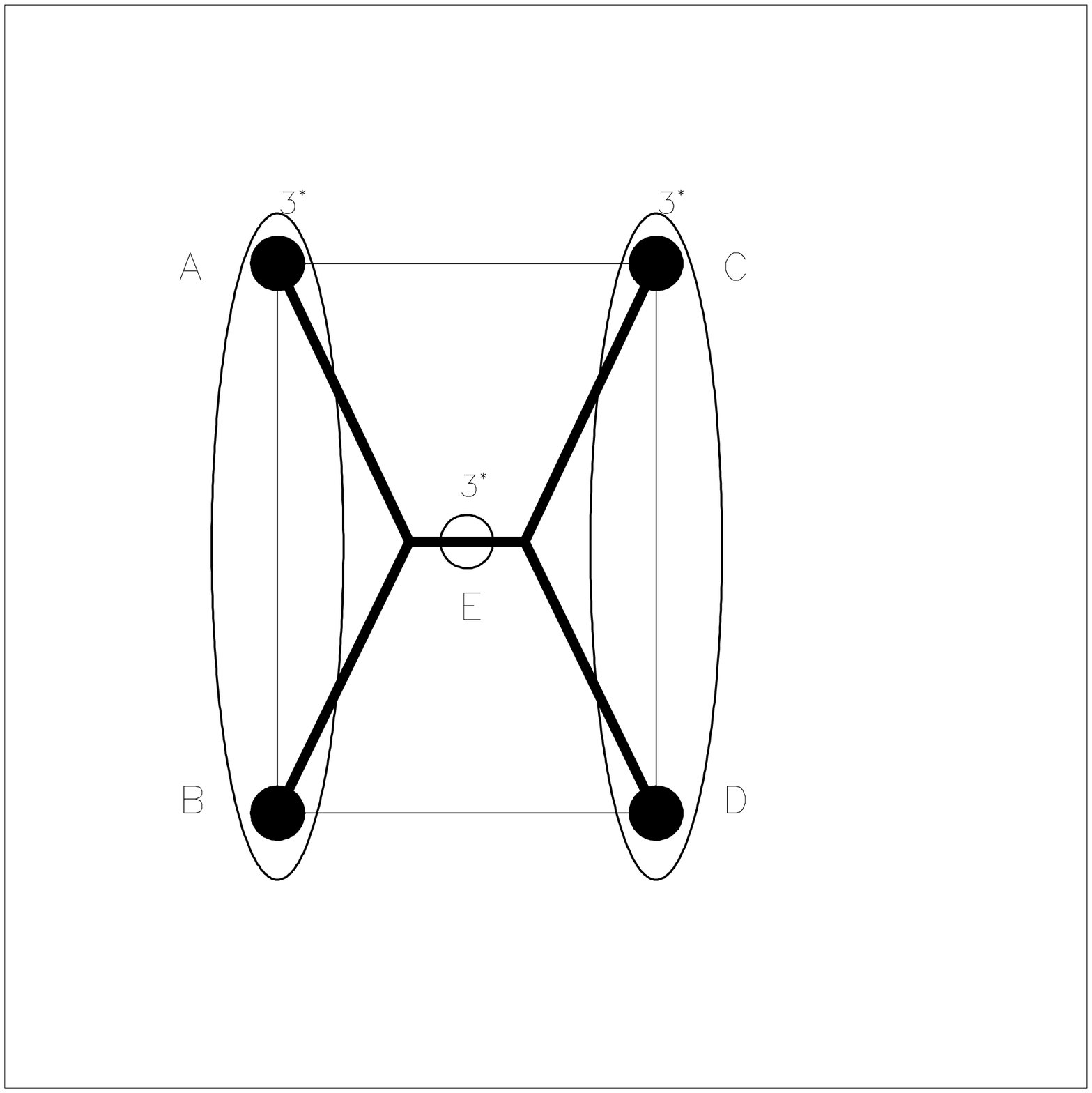}}
\hfill
\parbox{.32\textwidth}{\epsfysize=4.5cm \epsffile{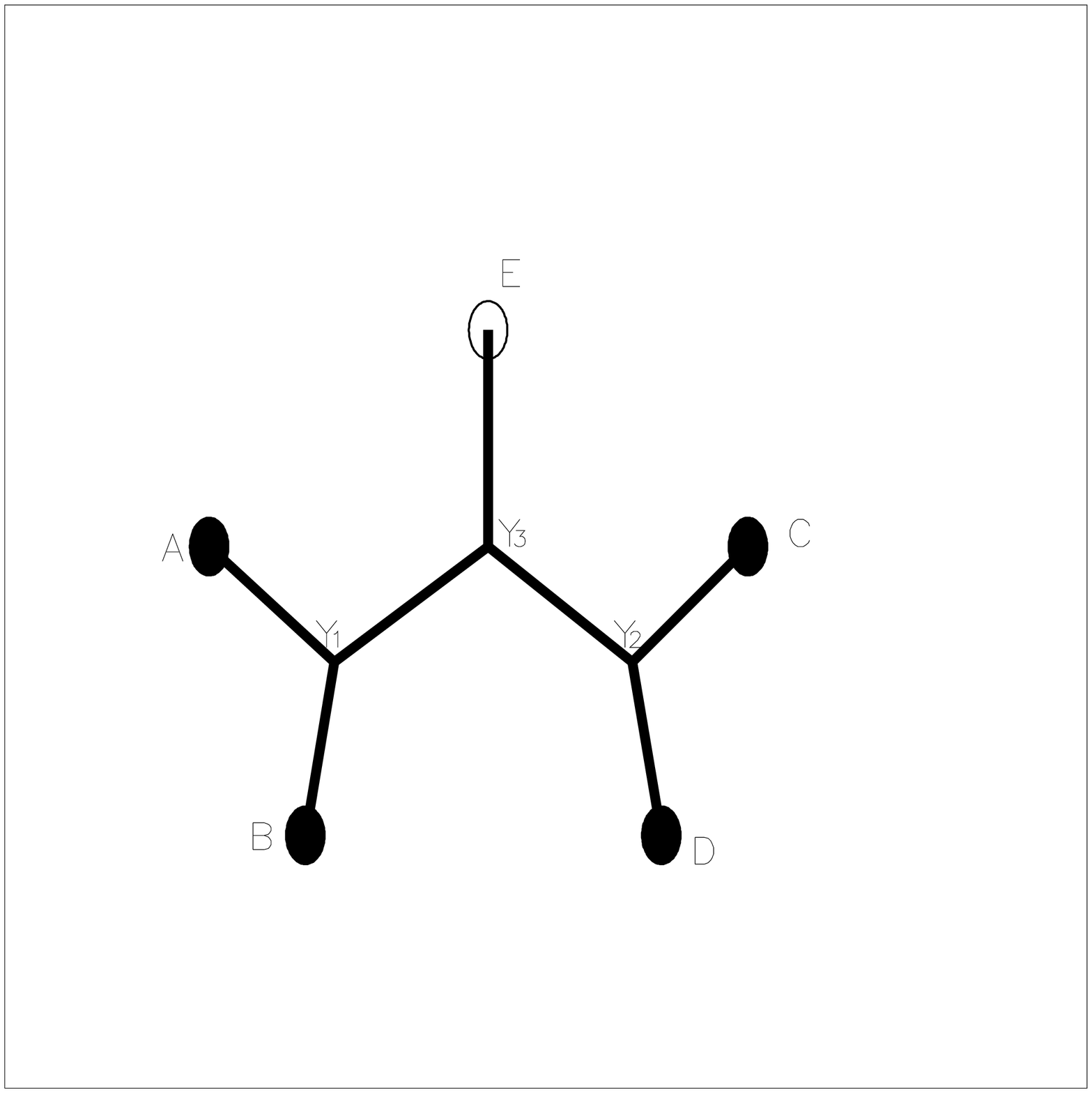}}
\hfill
\parbox{.32\textwidth}{\epsfysize=4.5cm\epsffile{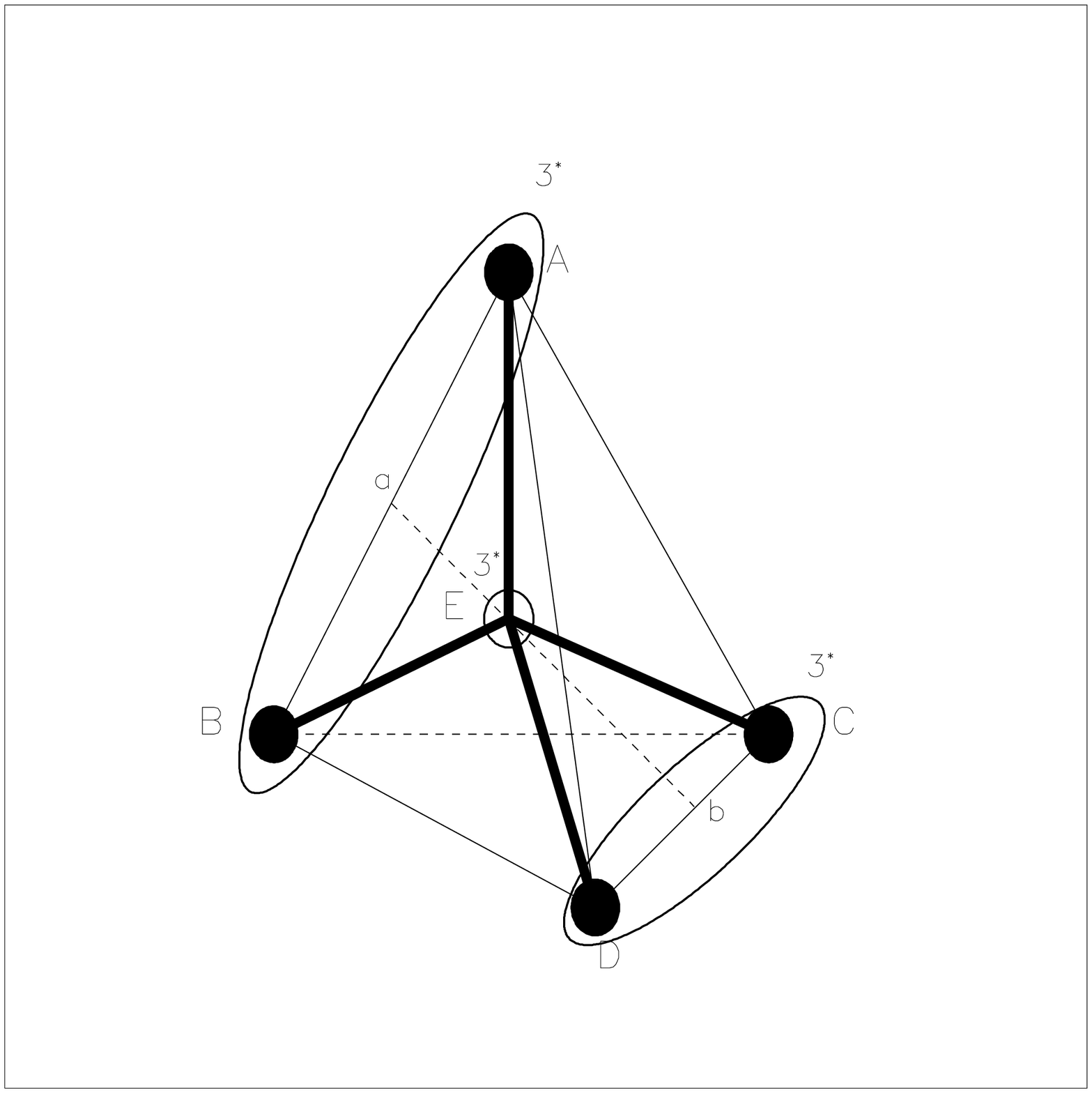}}
\hfill

\hspace{2cm}
\parbox{.25\textwidth}{\epsfysize=1.0cm (a)}
\hfill
\parbox{.26\textwidth}{\epsfysize=1.0cm (b)}
\hfill
\parbox{.20\textwidth}{\epsfysize=1.0cm (c)}
\hfill

\vspace{0.5cm}
\parbox{.33\textwidth}{\epsfysize=4.5cm\epsffile{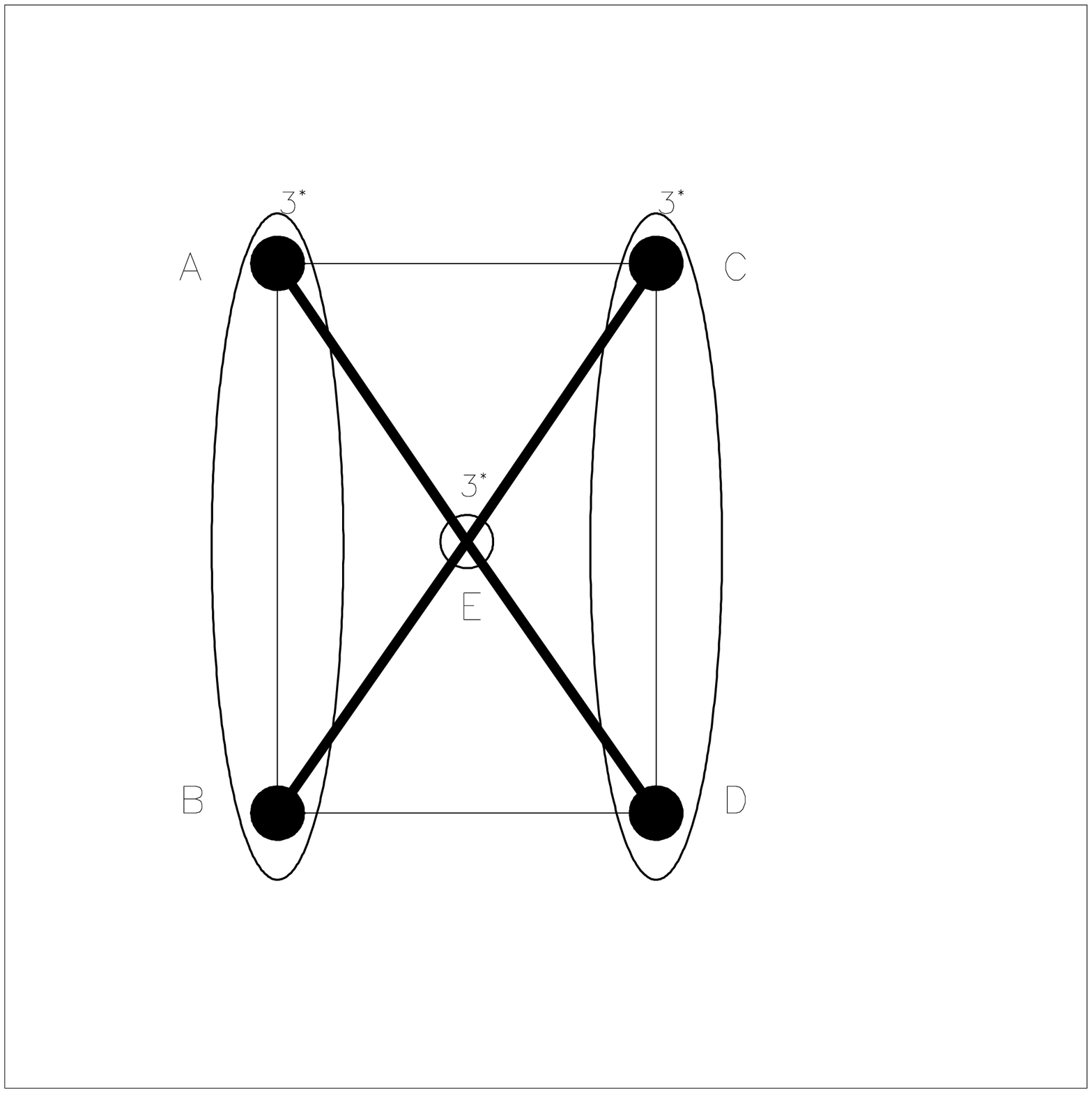}}
\hfill \hspace{5.1cm}
\parbox{.33\textwidth}{\epsfysize=4.5cm\epsffile{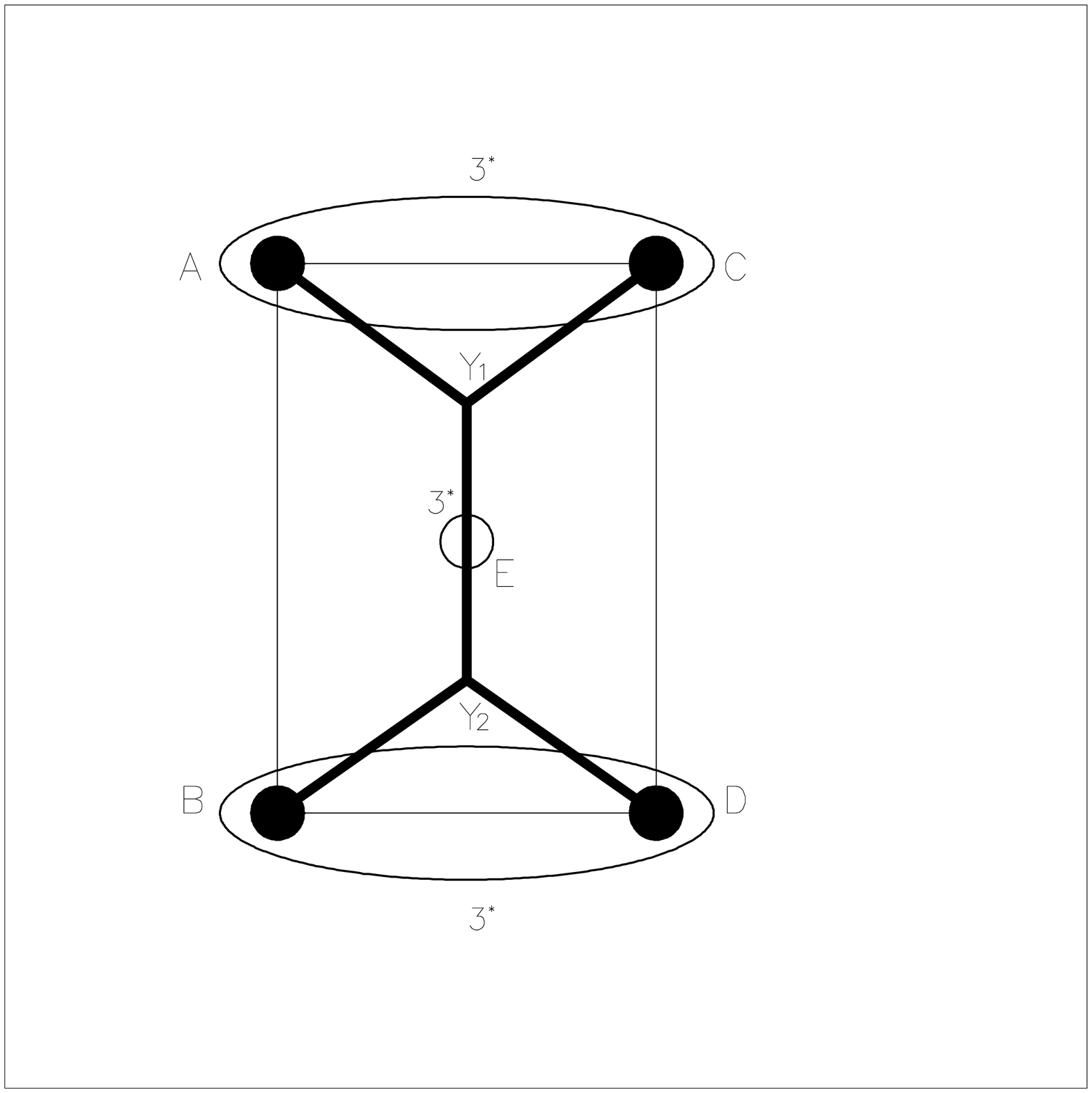}}

\hspace{2cm}
\parbox{.25\textwidth}{\epsfysize=1.0cm (d)}
\hfill \hspace{1.2cm}
\parbox{.19\textwidth}{\epsfysize=1.0cm (e)}
\hfill
\vglue 0.5cm
{\caption{\label{fig:conf} Various
configuration structures of pentaquarks.}}
\end{center}
\end{figure}

For a 5-quark system, the non-relativistic hamiltonian of the
system can generally be written as $H=T+V$. For the pictures
depicted in Fig.1, the four constituent quarks of the same mass
are residing at corners and the anti-strange quark occupies the
position of the center of mass of the four quarks. By the simple
picture, we further require that each configuration structure is
invariant, namely only the breathing mode is considered. That can
be understood from the classical angle, namely the relative
rotation or twist may cause an drastic energy increase (larger
tension), thus the main contribution to the energy of the ground
state comes from the simplest, but most significant mode, i.e. the
breathing mode. For convenience, the position of the anti-strange
quark is set to be the origin of the coordinate system. Let the
coordinate vectors of the four quarks be $\vec r_1$, $\vec r_2$, $\vec r_3$%
, and $\vec r_4 $, respectively, and obviously, $\vec r_1+ \vec
r_2+\vec r_3+\vec r_4 =0$. Then, the kinetic energy term of $H$
can be simplified as
\begin{eqnarray}
T =- \sum\limits_{j=1}^{4}
\frac{\hbar^2}{2m_j}\nabla^2_{\vec{r_j}}.
\end{eqnarray}
Obviously, in this simplified form, we ignore the oscillation of
the anti-strange quark about the center, i.e. keeping only the
breathing mode. The general form of the potential term of $H$ reads $%
V=V^{Coul}+V^{conf}+C_{5q}$, where $V^{Coul}$ is the
one-gluon-exchange(OGE) Coulomb potential which exists between
quark-quark and quark-anti-quark, $V^{conf}$ is the confinement
potential in the flux tube picture and $C_{5q}$ is a zero-point
energy. $V^{Coul}$ for the configurations in Figs.1(a), (c) and
(d) reads
\begin{eqnarray}
V^{Coul}=-A_{5q}[(\frac{1}{r_{12}}+\frac{1}{r_{34}})+\frac{1}{2}( \frac{1}{%
r_{15}} +\frac{1}{r_{25}}+\frac{1}{r_{35}}+\frac{1}{r_{45}})+\frac{1}{4}(%
\frac{1}{r_{13}} +\frac{1}{r_{14}}+\frac{1}{r_{23}}+\frac{1}{r_{24}})],
\end{eqnarray}
where $r_{ij}=|\vec{r_i}-\vec{r_j}|$. $A_{5q}$ is the effective
coefficient of the color Coulomb interaction, and its value can be
extracted from the lattice QCD result \cite{Oki}. For the
configuration in Fig.1(e),
\begin{eqnarray}
V^{Coul}=-A_{5q}[(\frac{1}{r_{13}}+\frac{1}{r_{24}})+\frac{1}{2}( \frac{1}{%
r_{15}} +\frac{1}{r_{25}}+\frac{1}{r_{35}}+\frac{1}{r_{45}})+\frac{1}{4}(%
\frac{1}{r_{12}} +\frac{1}{r_{14}}+\frac{1}{r_{32}}+\frac{1}{r_{34}})].
\end{eqnarray}
The confinement term has the form of
\begin{eqnarray}
V^{conf}=\sigma_{5q}L_{min},
\end{eqnarray}
where $\sigma_{5q}$ is the string tension and also extracted from
the lattice QCD \cite{Oki}.

The key point is $L_{min}$ which is the minimal total length of
the flux tube and has different values for different
configurations. As mentioned above,
only the breathing mode is considered in the calculation, we constrain that $%
|\vec r_1|=| \vec r_2|=|\vec r_3|=|\vec r_4|=r$, and the shape of
the configuration of the system remains invariant, namely the
system possesses only  one degree of freedom. In this
approximation, the expression of $L_{min}$ is greatly simplified
\begin{eqnarray}
L_{min}=\left\{
\begin{array}{ll}
4r & \mbox{~~~~~~~~~for Figs.1(a) and (c)} \\
\frac{2}{3}(\sqrt{3}+3\sqrt{2})r & \mbox{~~~~~~~~~for Fig.1(d)} \\
\frac{2}{3}(3+ \sqrt{6})r & \mbox{~~~~~~~~~for Fig.1(e)}
\end{array}
\right.
\end{eqnarray}
In our calculation, we ask all the quarks staying in the s-wave
and there is no any orbital excitation. The values of parameters
$A_{5q}$ and $\sigma_{5q} $ are borrowed from Okiharu's paper
\cite{Oki}
\begin{eqnarray*}
A_{5q}=0.1366,~~~~~~~~~~~~~\sigma_{5q}=0.165GeV^2.
\end{eqnarray*}
Constant $C_{5q}$ is determined in the following way: By
calculating the mass of nucleon with the same form of potential
structure, namely only the Coulomb interaction and the Y-shape
confining interaction are included, the Coulomb coefficient and
the string tension of the Y form confinement are taken to be
$A_{3q}=A_{5q}$  and $\sigma_{3q}=\sigma_{5q}$, respectively, we
extract $C_{3q}=-1.801GeV$. Therefore, we obtain
$C_{5q}=\frac{5}{3}C_{3q}=-3.002GeV$. Then there is no free
parameter in the potential anymore. Substituting the potentials
and kinetic piece into the Schr\"odinger equation and solving
\begin{eqnarray}
(T+V)\psi=E\psi
\end{eqnarray}
for the pentaquark system with $m_q=0.33GeV$ and $m_s=0.42GeV$
\cite{Mart}, we obtain the mass of $\Theta^+_5$ as
\begin{eqnarray}
M_{\Theta^+_5}=4m_q+m_s+E.
\end{eqnarray}
The resultant masses of $\Theta^+_5$ in various configurations are
tabulated in Table 1.
\begin{table}[hbt]
\begin{center}
\parbox{0.8\textwidth}{\caption{Resultant mass of $\Theta^+_5$. }}
\begin{tabular}{|c|c|c|c|c|}
\hline Configuration & Fig.1(a) & Fig.1(c) & Fig.1(d) & Fig.1(e)
\\ \hline $M_{\Theta^+_5}~(GeV)$ & 1.743 & 1.756 & 1.754 & 1.537 \\
\hline
\end{tabular}
\end{center}
\end{table}

It is seen that although the calculation is based on a very
simplified scenario, the obtained mass of the pentaquark is well
consistent with the LEPS' data \cite{Naka}. It is not simply an
coincidence or obtaining a result by fitting data,  the
plausibility of the approach is obvious.

Among the four configurations considered in the text, the mass of
the pentaqaurk with the structure shown in Fig 1(c), i.e. the
tetrahedron, has the largest value, which is about $0.216GeV$
above the central value of the mass measured by the LEPS
collaboration \cite {Naka}. As pointed by Okiharu \cite{Oki}, we
find that the masses of the twisted non-planar configuration (Fig.
1(c)) and the corresponding planar configuration (Fig. 1(d)) are
almost degenerate, and the planar one is slightly smaller. A
$0.003GeV$ difference comes from the difference of the color
Coulomb energy. From the point of view that energy should take the
minimal value, the planar structure is more favorite than others.

Indeed, one can note that the absolute value of the Coulomb energy
is inversely proportional to the distance between constituents,
whereas the confinement part is proportional to the total length
of the flux tubes, the two pieces compete with each other and
minimization of their sum results in a minimal binding energy.
When one investigates the configurations to determine which one
would be the most favorite one, he cannot simply decide by
comparing only $L_{min}$, but needs to consider the contributions
from both.

The total length of the flux tube with double Y-shape shown in
Fig.1(a) is shorter than that with the X-shape shown in Fig.1(d),
the resultant mass of the former configuration (Fig.1(a)) is
$0.011GeV$ smaller than that of the later (Fig.1(b)). If we keep
the shape of the configuration unchanged, the total length of the
flux tube in Fig.1(e) is even shorter. In addition, the concerned
color Coulomb energy is smaller compared with that resulted in by
the configuration Fig.1(a), the mass of the configuration in
Fig.1(e) becomes the smallest one among four calculated masses.
This means that the most favorite configuration is the one in
Fig.1(e). The obtained mass agrees with the LEPS' data value.

Moreover, when the masses of quarks increase, the direct
contribution from the constituent masses of quarks to
$M_{\Theta^+_5}$ increases, but the eigen energy of the system
decreases because then the kinetic energy contribution which is
inversely proportional to the constituent mass, takes a leading
role. Thus, there may exist an optimal value for the quark mass in
the physical region. We present dependence of the mass of
$\Theta^+_5$ on the quark mass in Table 2.
\begin{table}[hbt]
\begin{center}
\parbox{0.8\textwidth}{\caption{Resultant mass of $\Theta^+_5$ with
$m_{\bar{s}}=0.42GeV$ and different quark mass $m_q$.}}
\begin{tabular}{|c|c|c|c|c|c|c|}
\hline $m_q(MeV)$& 0.25 & 0.27 & 0.29 & 0.31 & 0.33 & 0.35
\\ \hline $M_{\Theta^+_5}~(GeV)$ & 1.534 & 1.524 & 1.523 & 1.527 & 1.539 &1.554 \\
\hline
\end{tabular}
\end{center}
\end{table}
From the result, one sees that when $m_q=0.29$ GeV, the minimal
$M_{\Theta^+_5}$ is $1.523$ GeV. Because the eigen-energy of the
system is independent of the mass of the anti-strange quark, if
$M_{\Theta_5^+}=1.54$ GeV is confirmed, at this energy scale
$m_{u(d)}=0.29$ GeV and $m_{\bar{s}}=0.44$ GeV would be the best
fit. In fact, from Table 2, one can note that as $m_q$ varies from
0.25 MeV to 0.35 MeV, the mass of $\Theta_5^+$ only changes by
1.3\%, namely is insensitive to $m_q$. But, we still can see that
there exists an optimal value for the mass of the light quark,
which is slightly smaller than that in the three quark system. In
our calculation, there is no fine tuning at all.

As mentioned above, in our model all the constituent quarks stay
at $S$-wave, and no orbital excitation is considered. The parity
of the pentaqurk is negative which comes from the parity of the
antiquark.

In summary, in the framework of the flux tube model where the
effective coefficient of the Coulomb interaction and the string
tension are extracted from the lattice QCD calculation, an
estimation based on the simplified semi-classical configurations
for the quark-antiquark disposal shows that $\Theta_5^+$ favors
the planar configuration, where two quarks at the ends of the
shorter edges of the rectangle form color-$\bar 3$ states, over
the tetrahedral configuration. And in this scenario, the parity of
$\Theta^+_5$ is negative. Certainly, as a simplified version, a
strong constraint is enforced on the possible motion mode of the
pentaquark system, i.e. only the breathing mode is accounted for.
Calculations based on more realistic structures of a five quark
system would be much more complex. A more sophisticated
calculation with the structure and the width of the pentaquak is
our under-going project.

This work is partially supported by the National Natural Science
Foundation of China under grant Nos. 10475089, 10435080, 10375090,
90103020 and CAS Knowledge Innovation Key-Project grant No.
KJCX2SWN02.


\begin{thebibliography}{9}
\bibitem{Naka} T.Nakano {\it {et al.}}, Phys. Rev. Lett. {\bf {91}},
012002(2003).

\bibitem{Zhu} S.L.Zhu, Int. J. Mod. Phys. {\bf {A19}}, 3439(2004) (and
references therein).

\bibitem{Isg} N.Isgur and J.Paton, Phys. Rev. {\bf {D31}}, 2910(1985).

\bibitem{Oki} F.Okiharu, H.Suganuma and T.Takahashi, hep-lat/0407001.

\bibitem{Alex} C.Alexandrou and G.Koutsou, Phys. Rev. {\bf {D71}},
014504(2005).

\bibitem{Song} X.C.Song and S.L.Zhu, Mod. Phys. lett. {\bf {A19}},
2791(2004).

\bibitem{Kana} Y.Kanada-En'yo, O.Marimatsu and T.Nishikawa, hep-ph/0410221.

\bibitem{Takah} T.T.Takahashi {\it {et al.}}, Phys.Rev.Lett., {\bf {86}}%
,18(2001), Phys.Rev., {\bf {D65}}, 114509(2002).

\bibitem{Mart} B.V.Martemyanov, C.Fuchs, Amand Faessler and
M.I.Krivorruchenko, Phys.Rev., {\bf {D71}}, 017502(2005).
\end{thebibliography}
\end{document}